**Main Manuscript for**

**Large role of anthropogenic climate change in driving smoke exposure across the western United States from 1992 to 2020**

Xu Feng[1,*], Loretta J. Mickley[1], Jed O. Kaplan[2], Makoto Kelp[3], Yang Li[4], Tianjia Liu[5]

[1]John A. Paulson School of Engineering and Applied Sciences, Harvard University, Cambridge, Massachusetts, USA

[2]Department of Earth, Energy, and Environment, University of Calgary, Calgary, Alberta, Canada

[3]Department of Earth System Sciences, Stanford University, Stanford, California, USA

[4]Department of Environmental Science, Baylor University, Waco, Texas, USA

[5]Department of Earth System Science, University of California, Irvine, California, USA

*Xu Feng

**Email:** xfeng@g.harvard.edu

**Author Contributions:** Paste the author contributions here.

**Competing Interest Statement:** The authors declare no competing interest.

**Classification:** Paste the major and minor classification here. Dual classifications are permitted, but cannot be within the same major classification.

**This PDF file includes:**

> Main Text
> Figures 1 to 4

**Abstract (not exceed 250 words)**

Wildfire activity has increased dramatically in the western United States (US) over the last three decades, having a significant impact on air quality and human health. However, quantifying the drivers of trends in wildfires and subsequent smoke exposure is challenging, as both natural variability and anthropogenic climate change play important roles. Here we devise an approach involving observed meteorology and vegetation and a range of models to determine the relative roles of anthropogenic climate change and natural variability in driving burned area across the western US. We also examine the influence of anthropogenic climate change on smoke exposure. We estimate that anthropogenic climate change accounts for 33-82% of observed total burned area, depending on the ecoregion, yielding 65% of total fire emissions on average across the western US from 1992 to 2020. In all ecoregions except Mediterranean California, anthropogenic climate change contributes to a greater percentage of burned area in lightning-caused wildfires than in human-caused wildfires. On average, anthropogenic climate change contributes 49% to smoke $PM_{2.5}$ concentrations in the western US from 1997 to 2020, and explains 58% of the increasing trend in smoke $PM_{2.5}$ from 2010 to 2020. We further find that populations in northern California, western Oregon, Washington, and parts of Idaho have experienced the greatest smoke exposure attributable to anthropogenic climate change in recent years. Our work highlights the significant role of anthropogenic climate change in degrading air quality in the western US and identifies those regions most vulnerable to wildfire smoke and thus adverse health impacts.

**Significance Statement (at least 50 words, not exceed 120 words)**

Wildfire activity has increased dramatically in the western United States over the last three decades, leading to a significant impact on air quality and human health. This study highlights the substantial role of anthropogenic climate change in driving these wildfires and the subsequent smoke exposure in the western US. Our results suggest that anthropogenic climate change contributes 33-82% to observed burned area in the western US from 1992 to 2020. Such climate change also accounts for nearly half of surface smoke $PM_{2.5}$ exposure from 1997 to 2020 and explains 58% of the increasing smoke trend from 2010 to 2020. Our study calls attention to the threat that wildfires have on smoke exposure and human health under a changing climate regime.

**Introduction**

Wildfire activity has increased dramatically across the western United States (US) during recent decades (1-3). The increasing magnitude and interannual variability of wildfires has been traced to a combination of anthropogenic climate change (ACC), natural variability, and human-caused ignitions (4-7). As a result, the burden of wildfire smoke–a complex mixture of carbonaceous aerosols, other types of fine particular matter ($PM_{2.5}$), and trace gases–has also increased, eroding multi-decadal efforts to reduce $PM_{2.5}$ concentrations in most western states (8, 9). Although $PM_{2.5}$ has significant impacts on human health and the economy (10-12), exploring the contributions of ACC and natural variability of human smoke exposure, not just fire activity, in the western US has been so far overlooked. The warming surface temperatures of ACC can dry out vegetation, increasing wildfire intensity or spread (4, 13); natural variability, driven by internal processes within the climate system, may also affect wildfire activity (14, 15). Improving our knowledge of the relative importance of these drivers of unhealthy smoke exposure could spur greater action to address ACC and would focus efforts of fire management on specific regions where wildfires are most likely to affect human health downwind.

Studies have found that climate change has been a major driver of fire weather and fuel aridity in the western US, explaining a large portion (50-70%) of the interannual variability and increasing trend of wildfire burned areas since the 1980s (2, 5, 16-18). Using fuel aridity as a proxy for burned area, Abatzoglou and Williams (4) suggested that ACC is responsible for doubling the trend in cumulative burned area in forested regions from 1984 to 2015. Beyond forested areas, other ecoregions have also experienced frequent wildfires in recent years, accounting for 60% of total burned areas across the western US from 1992 to 2020 (Fig. 1A and S1). However, the extent to which ACC and natural variability contribute to wildfires in these other ecoregions remains largely unquantified. Complicating the issue, however, is the evidence that the meteorological influence on wildfire activity varies by ecoregion (19). In addition, the western US is prone to both lightning-caused and human-caused fires, each with fundamentally different characteristics. Previous studies have shown that lightning-caused fires tend to be more intense and larger in size than fires related to human activity, and typically occur in summer and fall (June-October), while human-caused fires can occur at any time of the year (20, 21). These differences make it challenging to quantify the sensitivity of wildfires to different drivers.

Recent studies have explored how wildfires have affected present-day air quality and will likely degrade air quality in the future. For example, studies using air quality observations from monitoring stations and satellites have shown that wildfire smoke may now be the leading cause of poor air quality in the US, with especially high concentrations in the western US over the last decade (8, 9, 22). In addition, studies have estimated future wildfire emissions using statistical models or process-based, dynamic vegetation models (19, 23-26). By applying projected fire emissions to offline chemical transport models, researchers have quantified the impact of climate change on wildfire smoke exposure. In general, such studies have projected large increases in smoke exposure across the western US by 2050 or 2100, relative to the present-day (19, 23, 25-27). For example, over the Pacific Northwest, $PM_{2.5}$ pollution could double or triple during the fire season due to increased wildfire activity by 2100, depending on the greenhouse gas emission pathway (25). These previous studies, however, have not distinguished between the roles of ACC and natural variability in affecting wildfire smoke exposure in the recent past.

In this study, we investigate the extent to which ACC has already contributed to the increasing trends in wildfire smoke exposure across the western US from 1992 to 2020. First, we divide the western US into five ecoregions (28) and quantify the sensitivity of burned areas in these ecoregions to variations in climate variables. More specifically, we develop individual Gaussian Processing Regression (GPR) models to establish relationships between the observed burned areas of lightning-caused or human-caused fires and a suite of possible drivers, including meteorological variables and the normalized difference vegetation index (NDVI) in five ecoregions across the western US. We consider the fire activity through the entire year, not just during a typical fire season. The resulting timeseries of predicted area burned are referred to here as the "observed scenario." Second, we distinguish the signal of ACC from natural variability in the observed meteorology by using ensemble simulations from the sixth phase of the Coupled Model Intercomparison Project (CMIP6) (4, 5, 29, 30) (see Methods). We generate a natural climate scenario by first diagnosing the long-term ACC signals from the CMIP6 and then subtracting these signals from the observed meteorology. We apply the natural climate scenario to the GPR models for the lightning-caused and human-caused fires in each ecoregion, yielding timeseries of expected burned areas in a world unaffected by ACC over time. These predicted burned areas are then used to estimate natural fire emissions, following the framework of Global Fire Emission Database version 4.1 with small fires (GFED4.1s) (31). Finally, using the chemical transport model GEOS-Chem (32), we simulate the long-term surface smoke exposures across the western US, driving the model with fire emissions under observed and natural climate scenarios. The difference between these two scenarios represents the smoke exposure traced to ACC. In this way, we distinguish the effects of ACC from natural variability on smoke exposure and quantify with greater confidence the magnitude that this deleterious consequence of a warming climate has already had across the western US.

**Results**

**Historical trends of lightning-caused and human-caused fires and their drivers across the western US**

Our analysis of the Fire Program Analysis-Fire Occurrence Database (FPA-FOD) reveals that the contributions of lightning-caused and human-caused fires to total burned areas vary by ecoregion (see Methods and Supporting Information). During the last three decades, lightning-caused fires dominate fire activity in forested areas (Ecoregions 1 and 4) and shrublands (Ecoregion 3), accounting for 72% (Ecoregion 1), 75% (Ecoregion 4), and 62% (Ecoregion 3) of total burned areas (Fig. S1 and S2C). Human-caused fires contribute 86% to total burned areas in southern California (Ecoregion 2) and 55% to semi-arid prairies (Ecoregion 5) (Fig. S1 and S2D); these regions are mostly covered by crops and grassland.

Both lightning-caused and human-caused fires show large interannual variability from 1992 to 2020 (Fig. S1). We find that area burned of lightning-caused fires has significantly increased in Ecoregion 1 from 1992 to 2020, with a Theil-Sen estimated slope of 0.48 $\sigma$/decade ($\sigma$ represents the standard deviation of annual burned areas in the given ecoregion), and in Ecoregion 4 at 0.54 $\sigma$/decade. Significant increasing trends in human-caused fires also occur in Ecoregions 1, 3, and 5, with Theil-Sen slopes of 0.47 $\sigma$/decade (Ecoregion 1), 0.31 $\sigma$/decade (Ecoregion 3), and 0.23 $\sigma$/decade (Ecoregion 5). Other trends in burned area by lightning-caused or human-caused fires are not significant in their respective ecoregions. Over the entire western US, however, total burned areas of both lightning-caused and human-caused fires increase significantly by 0.52 $\sigma$/decade, with these increases dominated by fires occurring in the forested region of Ecoregion 1.

By comparing the GPR predictions with the FPA-FOD dataset, we find that meteorology and vegetation conditions can explain 36% to 85% of the interannual variability in total burned areas across the ecoregions in the western US (Fig. 1B). Here, the predicted total burned areas in each ecoregion are derived by summing the output from the GPR models for lightning-caused and human-caused fires in that ecoregion. The key predictors include daily maximum air temperature ($T_{max}$); moisture variables such as vapor pressure deficit (VPD), relative humidity (RH), specific humidity, and daily accumulated precipitation; and normalized difference vegetation index (NDVI). NDVI, measured daily by satellite, indicates the density and greenness of vegetation and can serve as an effective proxy for fuel loads.

For lightning-caused fires (Fig. S3), the GPR models accurately reproduce the mean burned areas in four ecoregions with normalized mean biases (NMBs) ranging from -4.2% to +2.6%, except for Ecoregion 2, where the mismatch is larger. The GPR models explain 79% (Ecoregion 1), 89% (Ecoregion 3), and 64% (Ecoregion 5) of the variance in burned areas, revealing that in these regions meteorology and vegetation control a large portion of interannual variability in lightning-caused fires. In Ecoregion 4, however, lightning-caused fires appear to be driven by other environmental or human factors, as the meteorology and vegetation variables chosen here account for only 44% of the interannual variability in burned areas. The GPR model performs poorly in capturing the variability of burned areas of lightning-caused fires in Ecoregion 2, where such fires are small and contribute only 14% to total burned areas.

For human-caused fires (Fig. S4), the GPR models slightly underestimate the mean burned areas in all regions with NMBs of -8% to -0.4%. As with lightning-caused fires, meteorology and vegetation also greatly control the extent of burned areas of human-caused fires, contributing 92% (Ecoregion 1), 64% (Ecoregion 4), and 85% (Ecoregion 5) to the interannual variability. In Ecoregions 2 and 3, GPR models explain ~40% of the variance in burned areas. This suggests that, more complex anthropogenic factors, such as the extent and management of grazing and agricultural land, have a strong influence on the burned area extent of human-caused fires. Despite these discrepancies in interannual variability, the small NMBs show that our GPR models can nonetheless capture the total amount of burned areas from both lightning-caused and human-caused fires over the last three decades.

**Contributions of anthropogenic climate change to burned areas across the western US**

Anthropogenic forcing leads to a warmer and drier climate across the western US with notable regional variations in air temperature, VPD, and precipitation (Fig. S5 and S6). We find that anthropogenic forcing results in increasing trends in the ensemble mean for $T_{max}$ of 0.48 ± 0.0063 K/decade (95% confidence interval, CI) and for VPD of 0.025 ± 0.0011 kPa/decade (95% CI) across the entire western US from 1992 to 2020 (Fig. S5A and S5B). These increasing trends for both $T_{max}$ and VPD, which we define as ACC signals, are significant across multiple CMIP6 models, with ranges of 0.06-0.76 K/decade for $T_{max}$ and 0.004-0.05 kPa/decade for VPD. The trend in the ensemble mean for daily accumulated precipitation over the entire western US for this time period is +0.025 ± 0.0033 mm/decade and is not statistically significant (Fig. S5C). Fig. S6 shows the spatial distributions of ACC signals for these variables averaged over 1992 to 2020 and the significance of the ACC trends in each grid cell. Results indicate that the increases in $T_{max}$ and VPD vary by ecoregion in the western US. The central and southwestern regions experience the greatest increases in $T_{max}$ and VPD, reaching 1.7 K and 0.18 kPa, respectively. However, anthropogenic forcing has contrasting effects on simulated precipitation in different regions, with these effects reaching statistical significance in most grid cells. Such forcing reduces average daily accumulated precipitation by 0.05 to 0.08 mm in the northern California and coastal area of Oregon, but slightly increases it by 0.02 to 0.05 mm in central areas.

Under the natural climate scenario, which excludes ACC signals, the burned areas predicted by GPR models account for only 35% of those in the FPA-FOD dataset in the whole western US from 1992 to 2020. Fig. 1B compares the total burned areas predicted under the natural climate scenario with those predicted using the observed climate scenario in five ecoregions over this time period. The differences in predicted burned areas under the two climate scenarios represent the contribution of ACC to the trends in total burned areas. In most years, ACC amplifies the predicted area burned, compared to that in the natural scenario. Our results suggest that ACC contributes 33-82% to total burned areas in the five ecoregions during the last three decades; this range reflects the different responses of lightning-caused and human-caused fires to ACC.

We further find that lightning-caused fires are more sensitive to ACC in most ecoregions, with ACC contributing 64-92% to the burned areas of lightning-caused fires (Fig. S3). The impacts of ACC on lightning-caused fires are higher in forest and shrubland ecoregions, with ACC contributions of 69% (Ecoregion 1), 79% (Ecoregion 3), and 92% (Ecoregion 4). Human-caused fires are somewhat less affected by ACC, with 35-64% of human-caused burned areas due to ACC in the five ecoregions (Fig. S4). ACC yields the lowest contributions (35%) to burned areas of human-caused fires in Ecoregion 2, the region that is most densely populated and characterized by a significant proportion of agricultural land.

Our results indicate that the dominant climatic factors driving burned areas vary by ecoregion (Fig. S7). In the most densely forested areas (Ecoregions 1 and 4), both lightning-caused and human-caused fires are most sensitive to increasing VPD; in grassland and shrub areas (Ecoregions 2, 3, and 5), both kinds of fires are most sensitive to increasing $T_{max}$. Furthermore, we also find different sensitivities of lightning-caused and human-caused fires to the identical ACC perturbations of VPD and $T_{max}$ (21).

**Contributions of anthropogenic climate change to fire emissions and smoke exposure across the western US**

We compare the time series of annual fire emissions of organic carbon (OC) under the observed and natural climate scenarios in the western US from 1997 to 2020 (see Methods). In the GFED4.1s inventory, annual fire emissions of organic carbon (OC) significantly increase by 11 Gg a$^{-1}$ over 1997 to 2020 in the western US, with cumulative emissions of 5.1×10$^3$ Gg. Under natural climate conditions, the increasing trend of annual OC fire emissions is only 3.5 Gg a$^{-1}$, with total emissions of 1.8×10$^3$ Gg summed over all years, or approximately one-third of the total in the observed scenario (Fig. 2A). The difference in annual OC fire emissions between GFED4.1s and the natural scenario represents the ACC contribution to these emissions, ~65% of total fire emissions. We find that for the whole western US, ACC explains ~68% of the increasing trend in OC emissions from 1997 to 2020. Across the five ecoregions, ACC contributes 26% to 76% to OC fire emissions (Fig. S8), proportional to the ACC contributions to total burned areas.

Our results indicate that large fire emissions driven by ACC occur in the forested regions along the Coastal Range, the Sierra Nevada, and the Cascade Range, and in northern Idaho, accounting for ~60% of current fire emissions in the western US from 1997 to 2020 (Fig. 2B). The mean ACC contributions range from 50-75% over the last two decades. The region where the fire emissions are most affected by ACC (~90%) lies in the Upper Gila Mountains, which stretch across Arizona and New Mexico (Fig. 2C). In southern California, where emissions are dominated by human-caused fires, the ACC contributions to these emissions are less than 50% (Fig. 2C). The rest of western US is covered mainly by savanna and shrubland, and we find that fire emissions account for only 2% of total emissions over the time period, due to low fuel loads and low emission factors for these land types.

To quantify the contributions of ACC and natural variability on long-term smoke exposure, we use the fire emissions under the two climate scenarios (observed and natural) as input to GEOS-Chem, a chemical transport model (see Methods). Table S1 shows the model configurations for three sensitivity experiments. The control experiment (CTL) and natural experiment (NAT) simulate $PM_{2.5}$ concentrations from both fire and non-fire sources under observed and natural climate scenarios. The magnitudes of non-fire sources (e.g., industry or dust mobilization) is the same in both scenarios. The background experiment (BKG) simulates only $PM_{2.5}$ concentrations from non-fire sources. Even though most smoke $PM_{2.5}$ is generated from August through October in the western US, we focus on annual means to better compare with non-fire $PM_{2.5}$, which may have a different seasonality (Fig. S9).

Based on the BKG simulation, we find that the background concentrations of $PM_{2.5}$ from non-fire sources decreases by 44% averaged over the western US from 1997 to 2020, mainly due to the reduction in emissions of air pollutants under the Clean Air Act (33) (Fig. S9A). In contrast, annual mean concentrations of total $PM_{2.5}$ plateau and then reverse over the most recent decade. Here the difference in annual mean concentrations between the CTL and the BKG experiments represents the smoke $PM_{2.5}$ contributed by wildfires under observed climate conditions (Fig. S9D). Our results show that the annual mean concentrations of smoke $PM_{2.5}$ averaged across the western US range from 0.2-4.0 μg m$^{-3}$ from 1997 to 2020, with an average 0.73 μg m$^{-3}$. Results further highlight a significant increasing trend in smoke $PM_{2.5}$ of 0.2 μg m$^{-3}$ a$^{-1}$ ($p < 0.05$) from 2010 to 2020. The contribution of smoke $PM_{2.5}$ to total $PM_{2.5}$ exhibits large interannual variability from 1997 to 2020, ranging from 5-62%, again with an increasing trend over the most recent decade (Fig. S10). During the extreme wildfire year of 2020, smoke $PM_{2.5}$ contributes 62% to total $PM_{2.5}$. We also find similar trends in smoke OA and smoke BC during this period (Fig. S9).

We find that ACC contributes nearly half (49%) to mean concentrations of smoke $PM_{2.5}$ over the western US from 1997 to 2020 (Fig. S9D), with annual mean concentrations ranging from 0.04 to 2.7 µg m$^{-3}$ averaged across the western US and a multi-year mean of 0.4 µg m$^{-3}$. These values are determined by differencing the concentrations under the observed and natural scenarios. Under the natural climate scenario, annual mean concentrations of smoke $PM_{2.5}$ range from 0.1-1.3 µg m$^{-3}$ averaged across the western US from 1997 to 2020. During the most recent decade (2010 to 2020), ACC accounts for 58% of the increasing trend in smoke $PM_{2.5}$, and the contribution of ACC to mean smoke $PM_{2.5}$ concentrations rises to 54%.

We next examine more closely the spatial distributions of the percent contributions of smoke $PM_{2.5}$ to total $PM_{2.5}$ in the western US over two time frames, from 1997 to 2009 and from 2010 to 2020 (Fig. 3). Our results show that wildfire smoke is especially abundant in northern California, Oregon, northern Idaho, and western Montana, where smoke $PM_{2.5}$ accounts for 25-80% of total $PM_{2.5}$ from 1997 to 2009 (Fig. 3A). Over the rest of the western US, the contributions of smoke $PM_{2.5}$ to total $PM_{2.5}$ are less than 10%. We find that both ACC and natural variability of climate drive the smoke $PM_{2.5}$ during this time frame, with 10-50% traceable to ACC (Fig. 3B). However, during the more recent period of 2010 to 2020, the impacts of smoke $PM_{2.5}$ expand to most of the states of northwestern US (Fig. 3D), and the contributions of smoke $PM_{2.5}$ to total $PM_{2.5}$ increase to 50-95% over northern California, Oregon, Washington, and Idaho. Our results indicate that this 2010-2020 enhancement of smoke $PM_{2.5}$ is mostly driven by ACC, accounting for 25-66% of total $PM_{2.5}$ exposure across these regions (Fig. 3E). The rest of the smoke $PM_{2.5}$ exposure is attributed to natural variability with percent contributions of 10-30% (Fig. 3F).

Since much of the western US is sparsely populated, it is instructive to examine population-weighted smoke exposure (8) (see Supporting Information). This metric allows policymakers and stakeholders to more clearly pinpoint where smoke has become a public health issue, especially smoke driven by ACC. To that end, we calculate the spatial distributions of gridded population-weighted smoke $PM_{2.5}$ concentrations and their attribution to natural variability and ACC for these representative years: 2000, 2005, 2010, and 2015 (Fig. S11), and 2020 (Fig. 4). Examining individual years illustrates how ACC may have shaped different fire seasons. We acknowledge, however, that accurate attribution of ACC to individual years requires more sophisticated modeling beyond the scope of this study. Here we find that from 2000 to 2010, annual population-weighted smoke $PM_{2.5}$ concentrations are 0.15-0.2 µg m$^{-3}$ over the entire western US, with only 20-40% of these concentrations attributed to ACC. However, the population-weighted concentrations increase to 0.61 and 5.2 µg m$^{-3}$ in 2015 and 2020, respectively, with ACC contributions rising to 44% and 62%. Our results indicate that populations in urban and suburban areas in California, western Oregon, and Washington are most vulnerable to the threat of wildfire smoke. In the extreme fire year 2020, we find that populated-weighted smoke $PM_{2.5}$ concentrations average 9.2 µg m$^{-3}$ over California and 18.3 µg m$^{-3}$ over Oregon. Also in that year, a hotspot of increased population-weighted smoke exposure appears along the Front Range urban corridor in Colorado. Compared to natural variability, ACC emerges as the dominant driver in 2020, contributing 60-70% to the population-weighted smoke exposure.

**Conclusions and Discussion**

By quantifying the ACC impact on wildfires and smoke exposure across the western US, our work builds on previous work documenting the influence of ACC on area burned in western US forests in recent decades (4, 7). First, we find that the influence of ACC on area burned largely depends on type of ecoregion and fire ignition source. The ACC contributions to burned areas in the forested regions (Ecoregions 1 and 4) range from 62-82% during 1992 to 2020, greater than the approximately 50% contribution reported by Abatzoglou and Williams (4) for similar forested areas, summed over 1984-2015. The increase in ACC contributions may reflect the influence of recent large fires in the western US. In relatively sparsely populated regions dominated by grassland or shrubs (Ecoregions 3 and 5), ACC accounts for 61-71% of burned areas. These ecoregions, including many national forests, parks, and wilderness areas, are primarily affected by lightning-caused fires that have been allowed to burn with less human intervention since 1970s (34). Thus, ACC has a significant impact in these ecoregions. Even in the densely populated Mediterranean California region (Ecoregion 2), where many anthropogenic factors may influence fire activity (27), we find that ACC accounts for 33% of area burned over 1992-2020. The lower ACC contribution in Ecoregion 2 is likely due to the high percentage of human-caused fires (over 70%), which are typically less intense, smaller in size, and can occur under higher fuel moisture conditions, compared to lightning-caused fires (21). Previous studies have also demonstrated that human ignitions triple the length of the wildfire season (20, 21). These findings imply that the occurrence of human-caused fires may be less dependent on the weather and fuel conditions than lightning-caused fires and so less sensitive to ACC.

Second, by diagnosing the ACC contribution to observed $PM_{2.5}$ trends in the western US from 1997 to 2020, our work extends previous studies that have identified the increasing impact of smoke $PM_{2.5}$ on air quality, such as Burke, *et al.* (9). Consistent with such studies, our results reveal an increase in smoke $PM_{2.5}$ around 2010, reversing a decreasing trend in total $PM_{2.5}$ across the western US. We find that 25-66% of total $PM_{2.5}$ exposure across the western US during 2010-2020 can be attributed to the ACC influence on fire emissions, with 58% of the increasing trend in smoke $PM_{2.5}$ linked to ACC. Relatively remote areas in Idaho and northern California experience the greatest ACC impacts on smoke $PM_{2.5}$ (Fig. 3). However, densely populated areas such as the Central Valley in California and the Willamette Valley in Oregon exhibit high levels of population-weighted smoke exposure (Fig. 4), making clear the large threat that ACC poses for human health.

We base our work on several assumptions. First, we assume that the interannual variability in burned areas in the western US is solely driven by meteorology and vegetation. However, other factors, such as changing land use and land management practices, as well as the history of fire suppression policies and prescribed burning, also contribute to the interannual variability, especially in areas outside national forests and parks (35, 36). Sources of human-caused wildfires include power lines, campfires, smoking, debris burning, equipment use, and arson (37), which are largely but not completely independent of specific meteorological and vegetation conditions. We also assume that the interannual variation in NDVI is entirely attributable to the natural variability of climate. Since fully dynamic land cover changes are not considered in the CMIP6 simulations used here, we apply the same time series of NDVI to both observed and natural climate scenarios in the regression models.

Exposure to ambient $PM_{2.5}$ ranks first among the leading causes of disease worldwide (38) and smoke $PM_{2.5}$ from wildfires thus represents a significant human health hazard (10, 11). While several studies have projected the influence of climate change on future smoke $PM_{2.5}$ in the western US (19, 23, 25), here we quantify the substantial effect that ACC has already had on smoke exposure in recent decades (1997-2020). Understanding the specific contributions of ACC versus natural variability in driving smoke exposure can help policymakers gain insights into where interventions may be most effective. For example, our study strengthens the case for aggressive and sustained land management in those regions where increasing wildfire activity generates the greatest smoke exposure downwind (39). By highlighting increased smoke exposure as a key consequence of a warming climate, this work could also spur development of stricter policies aimed at reducing greenhouse gas emissions.

**Methods**

**Wildfire burned area data**

We obtain the burned area data for lightning-caused fires and human-caused fires in the western US during 1992 to 2020 from the Fire Program Analysis-Fire Occurrence Database (FPA-FOD) (40). This database provides the discovery date, a point location, burned area, and fire cause of each wildfire record reported by a consortium of federal, state, and local organizations. We examine a total of 734,879 records of wildfires, each classified as having "Natural," "Human," and "Missing/not specified/undetermined" causes. Prescribed fires are not included in the FPA-FOD. We analyze the wildfire records labelled "Natural" and "Human," accounting for 33.5% and 58.3% of the total records. Records classified as having "Missing/not specified/undetermined" causes, about 8.2% of the total, are excluded in this study.

To validate the annual total burned area data from FPA-FOD, we use the satellite-derived burned area data from the Monitoring Trends in Burn Severity (MTBS) (41) (see Supporting Information). MTBS relies on data from Landsat and Sentinel-2. These data are analyzed using the differenced Normalized Burn Ratio algorithm (41) to generate the burned area data at a spatial resolution of 30 m. MTBS includes the burned area boundary, burn severity, ignition date, location, and fire types of each record since 1984. The three fire types are "Wildfire," "Prescribed Fire," and "Unknown." To compare with FPA-FOD, we use only the records classified as "Wildfire." MTBS contains records of only those wildfires with burned area greater than 1000 acres in western US, thus missing small fires (see Supporting Information).

To assess the interannual trends in total lightning-caused and human-caused fires, we use the Theil-Sen estimator (42, 43) to calculate the slopes and test the trend significance using the non-parameteric Mann-Kendall algorithm (44, 45) with a significance level of 0.05. To compare the spatial distribution of the two datasets, we match the burned areas to align with a 0.25° × 0.25° spatial resolution.

**Climate data and ACC signals from CMIP6 simulations**

Observed daily meteorological data from 1992 to 2020 are obtained from the gridMET dataset (46) and include maximum air temperature ($T_{max}$), vapor pressure deficit (VPD), minimum and maximum relative humidity ($RH_{min}$ and $RH_{max}$), specific humidity (SPH), accumulated precipitation (PRECIP), and wind direction (WD). The gridMET dataset has a spatial resolution of 1/24$^{th}$ degree (~4 km).

We rely on an ensemble of simulated meteorological output from CMIP6 to separate the long-term signals of ACC from natural variability. Use of an ensemble of climate results allows us to sample models with a range of strengths and biases (4, 5, 29, 30). We follow the method described in Chen*, et al.* (30), which defines the ACC signal as the multimodel ensemble mean of absolute differences between the 20-year moving averages of a particular variable (i.e., 1941-60, 1942-61, …, 2081-2100) and the mean value during the reference period (1921-1940). By relying on trends, rather than the absolute magnitudes of the CMIP6 variables, we avoid the biases inherent in these simulations. We construct an annual time series of each meteorological variable from the historical experiment (1920-2014) and the Shared Socioeconomic Pathway 5 – Representative Concentration Pathway 8.5 (SSP5-8.5) experiment of the available CMIP6 models for 2015-2020. The SSP5-8.5 does not differ significantly from the other pathways in this short, near-term period (47). We use the first member simulation of each CMIP6 model, driven by the same model configurations (r1i1p1f1) (48). Table S2 shows the availability of CMIP6 models for the meteorological variables selected for this study. We use near-surface air temperature and near-surface relative humidity to calculate VPD. The original data from CMIP6, which have varying spatial resolution, are all regridded to a common resolution of 1° × 1°. For the natural climate simulation, we subtract the ACC signal from observed gridMET data to obtain the timeseries of natural variability in each meteorological variable.

**Gaussian Processing Regression model**

We develop the regression relationships between potential drivers and observed burned areas of lightning-caused and human-caused fires in five ecoregions during 1992 to 2020 using the Gaussian Processing Regression (GPR), a machine learning algorithm based on a nonparametric Bayesian approach (49). The ecoregions are aggregated from the Level II Ecoregion data from the Environmental Protection Agency (Fig. 1A) (28). We choose the GPR model as it can model highly non-linear systems effectively and achieve good predictive performance with fewer data points than other methods (50). In this study, all predictors for the GPR models consist of the annual mean values of observed climate data from the gridMET dataset and the NDVI derived from the NOAA polar orbiting satellites (Advanced Very High Resolution Radiometer and Visible Infrared Imaging Radiometer Suite) (51, 52), while the target variables are the annual total burned areas of lightning-caused or human-caused fires obtained from FPA-FOD in each ecoregion. Use of observed data as predictors, rather than the CMIP6 output, avoids possible biases in the models and ensures that the interannual variability of the predictors matches that of the target variables. To select the predictors in each ecoregion, we build GPR models with different combinations of all variables and then select the one yielding the highest $R^2$ between the predicted and observed burned areas. Table S3 shows selected predictors for lightning-caused and human-caused burned areas in five ecoregions. We use 5-fold cross validation to evaluate the performance of the resulting GPR models. The resulting $R^2$ and root mean square errors (RMSEs) between the predictions from GPR models and FPA-FOD observations are also shown in Table S3.

To better understand the sensitivity of both lightning-caused and human-caused fires to each meteorological driver in the five ecoregions, we conduct a set of sensitivity experiments in which we exclude the ACC signals one by one from each of the observed meteorological variables applied to the GPR models. We then recalculate the burned areas using the resulting input datasets. Fig. S7 shows the individual and combined ACC contributions of meteorological variables to predicted burned areas of lightning-caused and human-caused fires in each ecoregion.

**Fire emissions for the observed and natural climate scenarios**

Total fire emission of a specific species in GFED4.1s is a function of dry matter burned, emission factors for different fire sources (grassland and savanna, woodland, deforestation and degradation, forest, agricultural waste burning, and peat fires), and the fractional contributions of each fire type in a grid cell. Dry matter burned is calculated by multiplying burned area by fuel consumption. Burned area data compiled for the GFED4.1s inventory are derived from the Moderate Resolution Imaging Spectroradiometer (MODIS) observations onboard the Terra and Aqua satellites (53); these data represent current burning levels under observed climate conditions. In GFED4.1s, fuel consumption statistics are estimated by the Carnegie-Ames-Stanford Approach (CASA) biogeochemical model (31). Emission factors for individual species (in units of g specie per kg dry matter burned are taken from published studies and vary with land cover (54-56). The GFED4.1s inventory, which has a spatial resolution of 0.25° × 0.25°, is available only from 1997 to 2020, and so all our smoke simulations cover this timeframe.

In this study, we rely on original data from the GFED4.1s as the fire emissions under the observed scenario, as the annual burned area data from the GFED4.1s inventory are in good agreement with the FPA-FOD data in each ecoregion (Fig. S12), which in turn are in good agreement with results from the GPR models. To calculate the expected fire emissions under the natural climate scenario, we use our predicted burned area data under this scenario following the GFED4.1s framework described above. We estimate dry matter burned under the natural climate scenario based on the relationship between dry matter burned and burned area in the GFED4.1s. Our approach assumes the same spatial distribution of area burned as in GFED4.1s, but scales dry matter burned. Fig. S13 illustrates the strong linear correlations ($R^2$ = 0.6-0.9) between the annual total dry matter burned and annual total burned areas from the GFED4.1s inventory in five ecoregions for 1997 to 2016. We scale the dry matter burned from the GFED4.1s inventory in each ecoregion by the ratios of predicted total burned areas under the natural climate scenario to those under the observed climate scenario, yielding the expected dry matter burned for the natural scenario.

**GEOS-Chem and setup of sensitivity experiments**

We use the chemical transport model GEOS-Chem (version 14.1.1) to simulate the long-term smoke $PM_{2.5}$ concentrations at the surface from 1997 to 2020. GEOS-Chem is a 3-D model of atmospheric composition driven by assimilated meteorological data from NASA Modern-Era Retrospective analysis for Research and Application version 2 (MERRA-2) (57). Here we use the nested-grid configuration for North America domain (40.625° W to 140° W, 10° N to 69.5° N) under both observed and natural climate scenarios. This approach allows us to focus on the ACC impact on just fire emissions and not on smoke transport. For computational efficiency, we apply the aerosol-only version of GEOS-Chem, which simulates $PM_{2.5}$ concentrations by considering the emissions of primary aerosol and aerosol precursors together with monthly mean oxidant fields archived from a GEOS-Chem benchmark simulation with a full chemical mechanism (58). The spatial resolution for the nested domain is 0.5° latitude × 0.625° longitude, with 47 vertical levels extended from the surface to 0.01 hPa. Initial and boundary conditions of chemical species concentrations are taken from global GEOS-Chem simulations at a resolution of 4° latitude × 5° longitude. The spin-up time for both global and nested-grid simulations is six months.

To attribute the contributions of ACC and natural variability on long-term smoke exposure, we conduct three sensitivity experiments with different fire emission inventories. To isolate the impacts of fire emissions under the different climate scenarios, we use the same assimilated meteorological data to drive the sensitivity experiments. The control experiment (CTL) relies on GFED4.1s emissions and attempts to reproduce the smoke exposure caused by observed wildfire conditions. For the natural experiment (NAT), we assume that the wildfires are under natural climate conditions, without being affected by ACC. The smoke exposure calculated in this experiment is due solely to natural wildfire emissions, constructed as described above. Fire emissions are emitted at the surface in the CTL and NAT experiments. Fire emissions are switched off in the background experiment (BKG). Sources of background $PM_{2.5}$ in the western US include dust, biogenic species, and anthropogenic emissions from the industrial, transportation, power plants, residential, and agricultural sectors. The main components of smoke $PM_{2.5}$ are organic aerosol (OA) and black carbon (BC), accounting for 80-90% of modeled smoke $PM_{2.5}$ from 1997 to 2020. OA concentrations in the model are estimated from OC by applying a mass ratio of OA to OC of 2.1 (59). The differences between the CTL and BKG simulated $PM_{2.5}$ concentrations represent smoke $PM_{2.5}$ concentrations under the observed climate scenario, while differences between the NAT and BKG simulated $PM_{2.5}$ concentrations represent smoke $PM_{2.5}$ under the natural climate scenario. The contribution of ACC to smoke $PM_{2.5}$ is estimated as the difference between the smoke $PM_{2.5}$ concentrations in the observed and natural scenarios. To evaluate the CTL simulation, we compare the modeled OC and BC with surface measurements from the Interagency Monitoring of Protected Visual Environment (IMPROVE) program (Fig. S14) (see Supporting Information).

**Acknowledgments**

This research has been supported by the Modeling, Analysis, Prediction, and Projection (MAPP) Program of the Climate Program Office in the National Oceanic and Atmospheric Administration (grant no. NA22OAR4310140).20

**Figures and Tables**

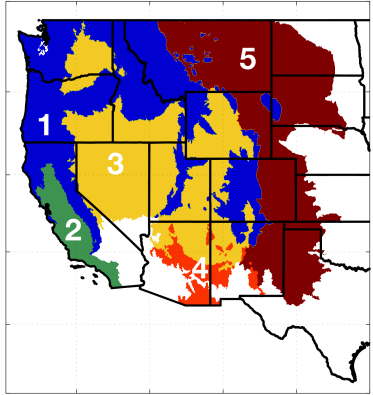

(a) Ecoregions in western US

| Ecoregion | North America Level II (EPA code) |
|---|---|
| 1 | Western Cordillera (6.2) |
|  | Marine west coast forest (7.1) |
| 2 | Mediterranean California (11.1) |
| 3 | Cold deserts (10.1) |
| 4 | Upper Gila Mountains (13.1) |
|  | Western Sierra Madre piedmont (12.1) |
| 5 | West-Central semi-arid prairies (9.3) |
|  | South-Central semi-arid prairies (9.4) |

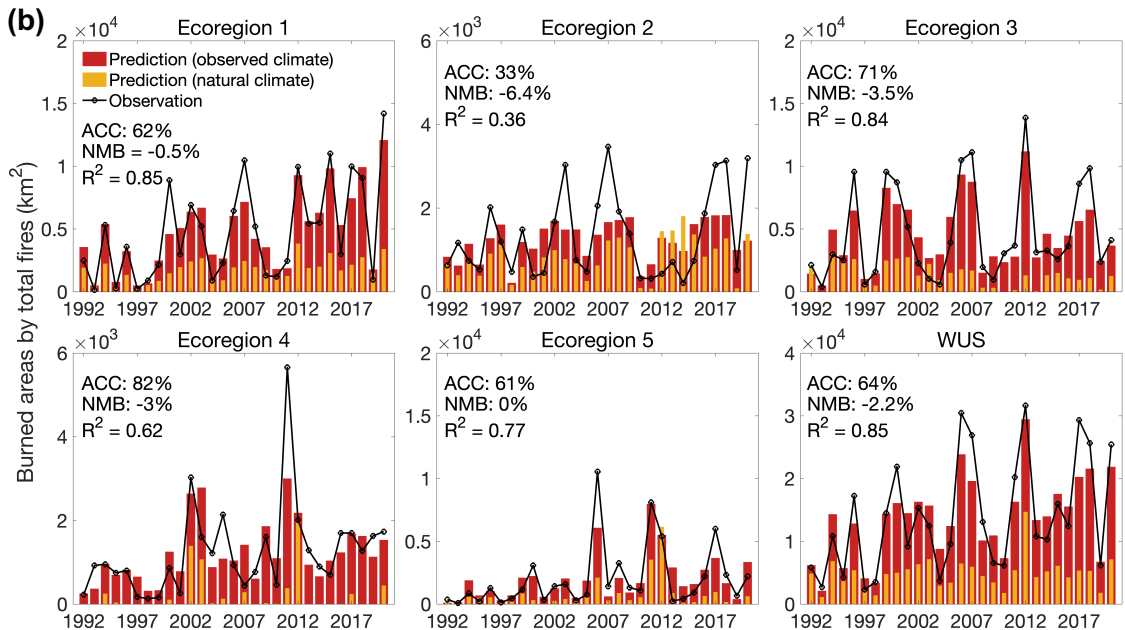

**Figure 1.** (a) Map of the five ecoregions in western US used in this study. Table describes the different ecosystems, which are aggregated from data from the Environmental Protection Agency. (b) Timeseries of annual total burned area from 1992-2020 in five ecoregions in the western US, as well as over the West as a whole. Black curves represent observed burned areas, including both lightning-caused and human-caused fires, from the FPA-FOD dataset. Colored bars show the predicted timeseries of annual burned area, calculated for observed (red) and "natural" (orange) conditions. The western US (WUS) represents the sum of burned area in the five ecoregions. Percent contributions of anthropogenic climate change (ACC) to the total predicted area burned are shown inset. The NMB and $R^2$ compare the predictions using observed climate conditions to the FPA-FOD dataset. Range of values on y-axes varies among panels.

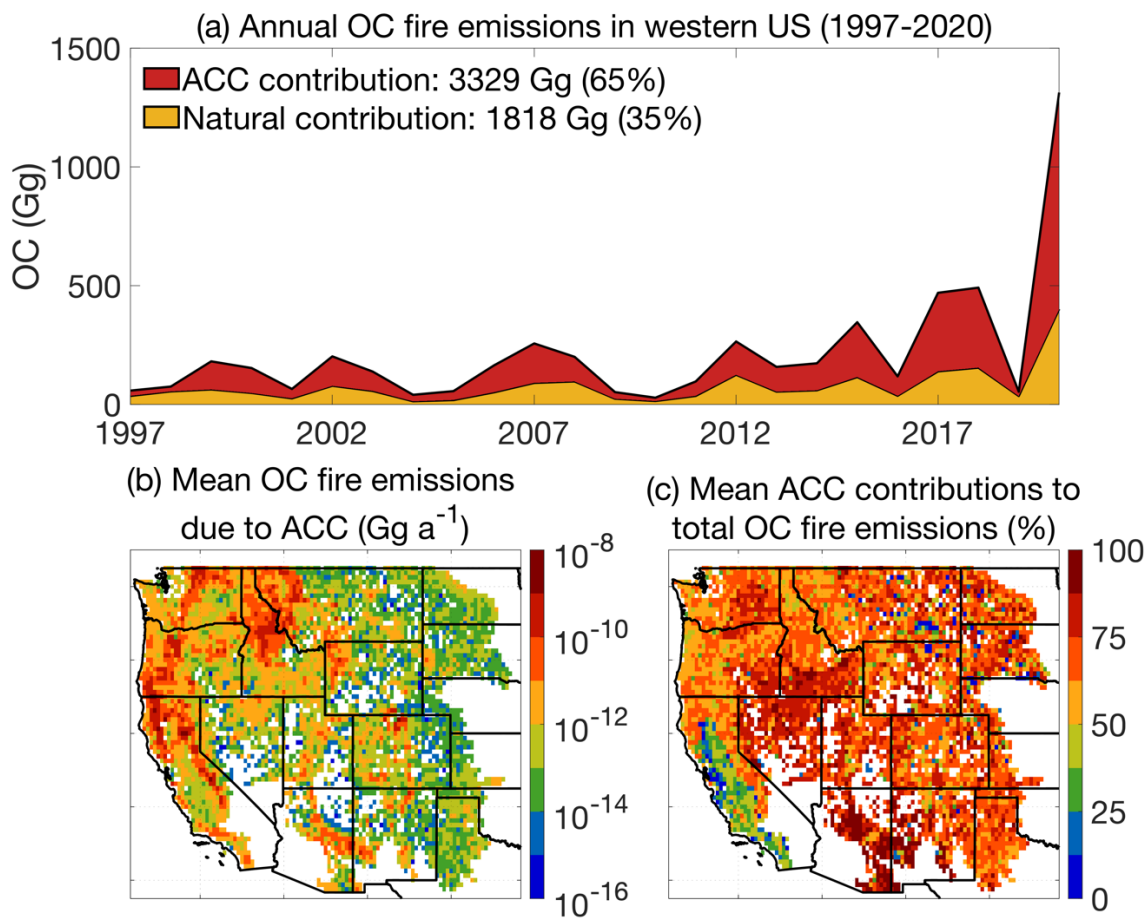

**Figure 2.** (a) Timeseries of annual total fire emissions of organic carbon (OC), the primary component of smoke $PM_{2.5}$, under the observed climate conditions (sum of red and yellow areas), as reported in the GFED4.1s inventory for the western US from 1997 to 2020. Also shown are the OC fire emissions under the natural climate scenario (yellow areas) and those emissions attributed to ACC (red areas). Values inset represent the OC fire emissions in Gg and the percent contributions to total OC fire emissions due to natural variability and to ACC, averaged over the time period. The bottom row shows the spatial distributions of (b) mean annual OC fire emissions due to ACC, averaged from 1997 to 2020, and (c) mean ACC percent contributions to total OC fire emissions in each grid cell.

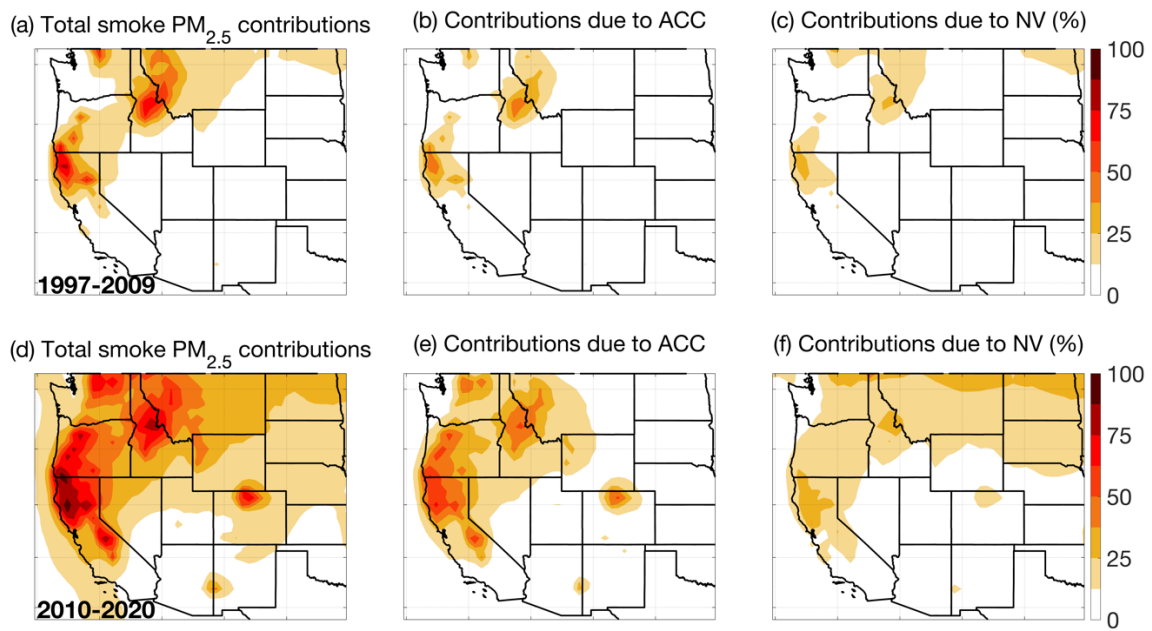

**Figure 3.** Calculated percent contributions of smoke PM$_{2.5}$ to total surface PM$_{2.5}$ for 1997 to 2009 (top row) and for 2010 to 2020 (bottom row). The panels show the spatial distributions of (a, d) contributions of all smoke PM$_{2.5}$ to total PM$_{2.5}$, (b, e) contributions of smoke PM$_{2.5}$ to total PM$_{2.5}$ due to ACC, and (c, f) contributions of smoke PM$_{2.5}$ to total PM$_{2.5}$ due to natural variability (NV).

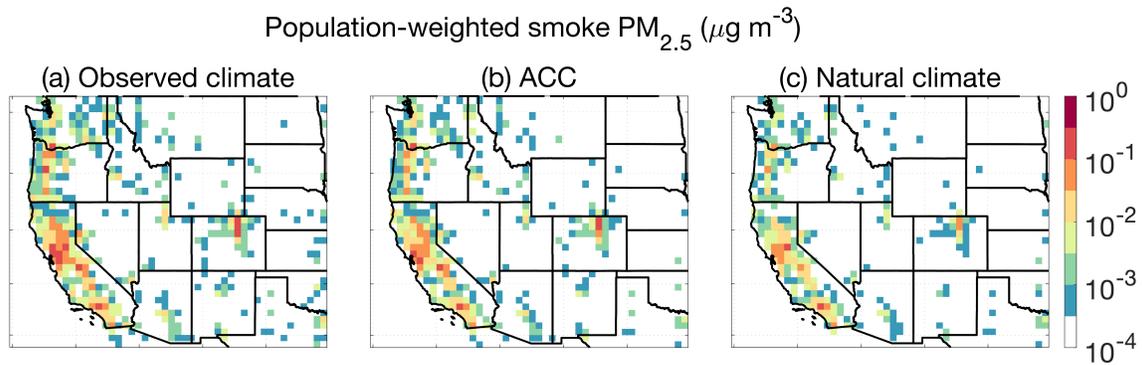

**Figure 4.** Spatial distributions of the annual average contributions to population-weighted smoke PM$_{2.5}$ exposure over the western US in 2020. The panels show the 2020 smoke exposures weighted by population in each grid cell, with hot spots revealing where smoke exposure affects the greatest numbers of people, compared with the rest of the West. The left and right columns show the smoke PM$_{2.5}$ exposures calculated by the CTL and NAT simulations. The center column shows the smoke PM$_{2.5}$ exposures attributed to ACC.